\documentclass[sigconf]{acmart}
\renewcommand\footnotetextcopyrightpermission[1]{} 
\usepackage{hyperref}
\usepackage{breakurl}

\usepackage[nameinlink, noabbrev, capitalize]{cleveref}
\usepackage{soul}

\begin{document}

\title[AI-Lab Framework for Computer Programming Courses]{Innovating Computer Programming Pedagogy: The AI-Lab Framework for Generative AI Adoption.}
\date{August 2023}

\author{Ethan Dickey}
\authornote{Both authors contributed equally to this research.}
\affiliation{%
  \institution{Purdue University}
  \streetaddress{Department of Computer Science}
  \city{West Lafayette}
  \state{Indiana}
  \country{USA}
  \postcode{47907}
  }
\email{dickeye@purdue.edu}
\orcid{0009-0007-3706-5253}

\author{Andres Bejarano}
\authornotemark[1]
\affiliation{%
  \institution{Purdue University}
  \streetaddress{Department of Computer Science}
  \city{West Lafayette}
  \state{Indiana}
  \country{USA}
  \postcode{47907}
  }
\email{abejara@purdue.edu}
\orcid{0000-0003-2611-2855}

\author{Chirayu Garg}
\affiliation{%
  \institution{Purdue University}
  \streetaddress{Department of Computer Science}
  \city{West Lafayette}
  \state{Indiana}
  \country{USA}
  \postcode{47907}
  }
\email{garg104@purdue.edu}
\orcid{0009-0009-6335-4686}

\renewcommand{\shortauthors}{Dickey, Bejarano, and Garg}

\begin{abstract}

  Over the last year, the ascent of Generative AI (GenAI) has raised concerns about its impact on core skill development, such as problem-solving and algorithmic thinking, in Computer Science students. Preliminary anonymous surveys show that at least 48.5\% of our students use GenAI for homework. With the proliferation of these tools, the academic community must contemplate the appropriate role of these tools in education. Neglecting this might culminate in a phenomenon we term the ``Junior-Year Wall,'' where students struggle in advanced courses due to prior over-dependence on GenAI. Instead of discouraging GenAI use, which may unintentionally foster covert usage, our research seeks to answer: ``How can educators guide students' interactions with GenAI to preserve core skill development during their foundational academic years?''
  
  We introduce ``AI-Lab,'' a pedagogical framework for guiding students in effectively leveraging GenAI within core collegiate programming courses. This framework accentuates GenAI's benefits and potential as a pedagogical instrument. By identifying and rectifying GenAI's errors, students enrich their learning process. Moreover, AI-Lab presents opportunities to use GenAI for tailored support such as topic introductions, detailed examples, corner case identification, rephrased explanations, and debugging assistance. Importantly, the framework highlights the risks of GenAI over-dependence, aiming to intrinsically motivate students towards balanced usage. This approach is premised on the idea that mere warnings of GenAI's potential failures may be misconstrued as instructional shortcomings rather than genuine tool limitations.

  Additionally, AI-Lab offers strategies for formulating prompts to elicit high-quality GenAI responses. For educators, AI-Lab provides mechanisms to explore students' perceptions of GenAI's role in their learning experience.
\end{abstract}



\maketitle
\pagestyle{plain}


\section{Introduction}

In recent years, the pervasive presence of GenAI tools across the internet has transformed the educational landscape. Integrating AI tools into educational environments has garnered substantial attention, yielding invaluable insights and recommendations for its implementation \cite{Dimitris16, Chassignol18, Chen20, holmes2022}. Nonetheless, the rapid emergence of prominent GenAI tools, including GitHub Copilot\footnote{\url{https://github.com/features/copilot}} in June 2021 and ChatGPT 3.5\footnote{\url{https://chat.openai.com/}} in November 2022, has prompted educators from diverse disciplines to raise apprehensions regarding students' utilization of these tools for academic advancement \cite{Terry23, Nolan23, PBS23}.

Since their emergence, GenAI tools have captured the attention of educators within the realm of computer science. Notably, Puryear and Sprint \cite{Puryear22} delved into the impact of \textit{Artificial Intelligence-Driven Development Environments (AIDEs)}, exemplified by tools like GitHub Copilot, within introductory Computer Science and Data Science courses. The authors undertook a comprehensive examination, revealing that Copilot could generate solutions for assignments across these courses, achieving scores ranging from 68\% to 95\%. However, the authors raised an overarching concern revolving around low code similarity scores when assessing potential plagiarism using tools such as MOSS\footnote{\url{https://theory.stanford.edu/~aiken/moss/}}. Their insights culminated in an encouraging directive, advocating for \textit{``computer science educators to become familiar with how AIDEs work and begin to design their coursework and development workflow to incorporate them''}.

Wermelinger's investigation \cite{Wermelinger23} and Moradi \textit{et al.}'s parallel study \cite{Moradi23} have both delved into the implications of GenAI on computing education and software engineering, respectively. Wermelinger assessed the capabilities of GitHub Copilot in comparison to OpenAI Codex's\footnote{\url{https://openai.com/blog/openai-codex}} Davinci model\footnote{\url{https://help.openai.com/en/articles/6195637-getting-started-with-codex}} for addressing CS1 coursework challenges, noting that while Copilot's performance did not rival Davinci's \cite{Finnie22}, it \textit{``does generate code (and with some editing, tests, and explanations) that could have been written by humans.''} On the other hand, Moradi \textit{et al.} explored Copilot's efficacy in the Software Engineering domain, particularly its precision in generating code for fundamental algorithms and data structures. Their findings highlighted Copilot's struggles in synthesizing multiple methods into a unified functional solution and emphasized that \textit{``an expert developer is still required to detect and filter its buggy or non-optimal solutions.''} Both studies underscore the importance of human expertise and urge educators to help students understand such tools' potential benefits and inherent limitations.



Since November 2022, ChatGPT 3.5 has emerged as the centerpiece of concern for educators grappling with GenAI tools. Its performance trajectory prompted the developers of OpenAI Codex to deprecate it surprisingly quickly\footnote{\url{https://platform.openai.com/docs/guides/code}}. Within a few months of the ChatGPT 3.5 release, instructors overseeing computer programming courses noted its proficiency in accurately solving numerous problems found within core Computer Science curricula \cite{Becker23, Yilmaz23, Ouh23, LauGuo23, SavelkaEtAl23, Perkel23, Nolan23, Hellas23}. Nevertheless, the tool's effectiveness dwindles as the coursework assumes more intricate dimensions, often resulting in inaccuracy and hallucinations. To mitigate these limitations, users can leverage fundamental prompt engineering techniques \cite{Giray2023, Hellas23}.

Evidently, the pace of progress in the tech industry outpaces academia. The release of ChatGPT 4 in March 2023 was highlighted by the remarkable improvements over its predecessor \cite{openai2023gpt4}. As expected, ChatGPT 4 garnered immediate attention from educators. They promptly raised concerns about its impressive capabilities and advocated reevaluating assessment paradigms to accommodate these evolving tools effectively \cite{Dobslaw23}. With the public release of the upgraded ChatGPT, educators now face a dilemma: should they actively promote student use of these tools or discourage their adoption? \cite{LauGuo23}.

The disruption brought about by GenAI tools has resonated deeply within the realm of Software Development. A simple search for \textit{``coding''} on the webpage \textit{There is an AI for that}\footnote{\url{https://theresanaiforthat.com/}} by August 2023 yields over 150 AI tools tailored for coding-related tasks, encompassing aspects like generation, completion, quality enhancement, productivity augmentation, and documentation. Among these, certain tools specifically designed for coding education have shown promise in enhancing the learning experience \cite{liffiton2023codehelp}. A subset of the tools are freely accessible to the public, seamlessly integrating with conventional \textit{Integrated Development Environments (IDEs)}. As we have underscored earlier, the enduring presence of these tools is indisputable, with current students likely to encounter their widespread application within the tech industry. Mindful of this level of accessibility and the evolving academic landscape in computing education, we have embraced these tools within our courses. Our approach entails equipping students with the prowess to harness these resources effectively while nurturing their foundational skills --such as problem-solving, algorithmic thinking, and debugging—- essential for holistic professional growth.

To address this objective, we pose the fundamental question: \textbf{``How can educators guide students' interactions with GenAI to preserve core skill development during their foundational academic years?''} Existing guidance within this domain tends to adopt a broader academic perspective, focusing on classroom utilization. Strategies involve empowering students to critically assess the accuracy and efficacy of GenAI outcomes alongside fostering dialogues about ethical considerations \cite{Gutierrez23, Guerriero23, Su23}. In the context of computing education, Bull \textit{et al.} propose a pedagogical strategy that acquaints students with how professional software developers harness GenAI within the tech industry. While emphasizing the imperative for students to emerge as adept problem-solvers by graduation, this approach underscores the utilization of scaffolding and fading techniques, iterative assignment modifications, and a gradual immersion trajectory \cite{Bull23}. Remarkably, up until the submission date of this paper, we have encountered no preexisting framework specifically aiding educators in incorporating GenAI tools within computer programming courses. A conspicuous gap among the cited frameworks is the need for delineated instructor steps for effectively integrating students into using GenAI tools while concurrently fostering the development of their core skill set.

This paper introduces the \textit{AI-Lab} framework, designed to seamlessly incorporate GenAI tools into computer programming courses. The framework functions as a structured environment where students can comprehensively understand the advantages and limitations inherent in GenAI utilization. Simultaneously, it furnishes instructors with valuable insights into students' viewpoints regarding using these tools within the present and upcoming courses. AI-Lab features prompt guidelines designed to optimize usage and facilitate effective communication. Additionally, the framework offers an interactive lecture dynamic, empowering instructors to engage students in substantive discussions surrounding course content while illuminating the boundaries and constraints of GenAI.

\section{The Junior-Year Wall Problem}

In the evolving landscape of Computer Science education, we identify a potential challenge that may arise as students transition into their junior year. Termed the \textit{Junior–Year Wall}, this concept encapsulates the scenario where students, having learned to rely on GenAI tools during foundational courses, find themselves facing difficulties when confronted with advanced topics, due to an incomplete grasp of core skills and the sudden lack of GenAI's ability to provide assistance.

\subsection{The Role and Implications of GenAI Tools: A Rational Exploration}
GenAI tools have undeniably revolutionized the way students approach foundational Computer Science courses. While there is no concrete evidence as of this submission date directly linking the use of these tools to the Junior–Year Wall problem, it is rationally conceivable. If students navigate their initial courses heavily aided by GenAI, there's a plausible risk that they might not fully internalize the content. As they advance to more complex subjects, where the efficacy of GenAI tools becomes significantly reduced, they could be confronted with significant challenges stemming from gaps in foundational knowledge acquired in earlier courses.

\subsection{Proactive Pedagogy: Preparing for Potential Challenges}
As we navigate the potential challenges posed by GenAI tools, it's imperative to consider how we can proactively integrate these tools into the curriculum without compromising the depth and quality of student learning. Our discourse is not a critique of GenAI tools. In fact, when judiciously incorporated into the curriculum, these tools can provide invaluable support, enabling students to address coding errors, gain diverse perspectives on concepts, and engage in hands-on practice. Such integration can reflect the real-world dynamics of software developers interfacing with AI tools in the industry.

Our stance is informed by two primary considerations:
\begin{enumerate}
    \item Educating students on the adept use of GenAI tools, while emphasizing the importance of mastering foundational concepts and illuminating their limitations, can serve as a robust deterrent against potential over-reliance. This proactive approach resonates with established principles of motivation theory \cite{howard_motivation, NASEM_motivation, ryan_motivation}.
    \item By anticipating potential educational challenges, we can tailor our pedagogical strategies to ensure comprehensive learning experiences, championing principles of equity and fairness.
\end{enumerate}

\subsection{Conclusion and Forward Momentum}
The Junior–Year Wall Problem, while not yet empirically validated, represents a rationally anticipated challenge in Computer Science education. Our role as educators is not merely to react to challenges but to anticipate and prepare for them. By recognizing the potential implications of the evolving educational landscape, we can ensure that our students are not only well-equipped with tools but also fortified with the critical thinking and foundational skills essential for their academic journey.

\section{AI-Lab Framework}

We introduce the AI-Lab framework as an integral course activity designed to educate students on using GenAI tools within their educational journey. The primary objectives of the AI Lab are two-fold: (1) to highlight the fallibility of AI, specifically its potential to produce incorrect outcomes, and (2) to emphasize the crucial link between robust skill development and the ability to tackle advanced tasks beyond the reach of AI assistance. The lab's implementation spans a range of programming concepts. Initially, students are encouraged to leverage GenAI for addressing questions and enhancing material comprehension. As the lab unfolds, progressively intricate queries are posed, strategically designed to surpass GenAI's problem-solving capabilities and consequently render erroneous responses. Significantly, this phase underscores the imperative of fundamental skill acquisition; any over-reliance on GenAI hampers students' capacity to accomplish tasks beyond its purview. The culmination of the lab incorporates a concise discourse wherein students reflect on their GenAI encounters, fostering critical reflection on AI's limitations and their learning experiences.

The effective integration of GenAI tools into computer programming courses can promote equity in the educational landscape. Equity stands as a pivotal catalyst driving our framework's trajectory. By equipping students with the skillset to proficiently navigate GenAI tools, we address the disparity faced by students who lack the privilege of pre-collegiate exposure to such resources. This proactive approach diminishes the ability gap, ensuring students enter higher education more equitably.

\subsection{Pre-Lab}

As with any course activity, the instructor holds the crucial responsibility of defining the scope of the AI-Lab. This initial task, conducted offline, necessitates a comprehensive formulation that includes clear objectives, goals, and desired learning outcomes. Additionally, the instructor must be diligent in selecting a topic that strategically challenges GenAI's efficacy. A paramount objective of this framework remains highlighting the pitfalls of unreflective GenAI reliance, which impede skill development. Hence, instructors must meticulously curate a topic that inherently heightens the likelihood of GenAI producing erroneous outcomes. For instance, the realm of coding solutions, well-established for ChatGPT's limitations, provides an illustrative case \cite{kabir23}. Our experience with GenAI tools has revealed that concepts involving pointer manipulation and concurrently managing multiple data structures within a singular algorithm tend to confound the tool's performance.

The pre-lab phase consists of three activities that students must complete prior to the ensuing in-person lab session (to be elaborated on in the following subsection). The underlying objective of the pre-lab phase is to provide students with an initial exposure to the chosen GenAI tool, fostering a sense of familiarity and comfort in its application. The three activities are the following:

\subsubsection{First Activity: Perceptions Survey.} 
Students must complete a survey that gauges their perceptions regarding the use of GenAI tools. The survey seeks to capture students' firsthand experience with GenAI tools and their observations of how their peers use them. This preliminary survey is crucial, furnishing instructors with valuable insights into students' perspectives on GenAI tool utilization within an academic context. We provide the pre-lab survey questions in an appendix in the full version of the paper.

\subsubsection{Second Activity: Familiarity with the tool.} Students must create an account on the chosen GenAI tool and acquaint themselves with it. This process involves providing students with detailed instructions, including illustrative prompts they can directly copy and paste. These example prompts serve as a means for students to engage with the tool's functionality firsthand.

Subsequently, students embark on their tool exploration using self-generated prompts. With chat-based GenAI tools, students can begin with a broad query, such as \textit{``Tell me about yourself,''} and continue the interaction by posing elementary questions. This dynamic mirrors human conversation, allowing students to engage with the tool as if conversing with another human. Other potential prompts are:
\begin{itemize}
    \item What are some of the places to visit in [your city]?
    \item Does [your university] actually have anything fun to do?
    \item What is a good recipe for [some dessert]?
    \item Help me plan a trip to [a place you want to visit].
\end{itemize}

Upon establishing proficiency with these introductory inquiries, students are encouraged to transition toward queries that align with their domain of study in computer science. We suggest providing prompts relevant to topics covered in the course curriculum. These prompts serve as a mechanism for students to assess the accuracy and comprehensiveness of the tool's responses. It is imperative to note that these provided prompts are deliberately concise, comprising no more than a single sentence. Students are empowered to pose questions about topics covered in the course, leveraging the tool to revisit, reinforce, or seek further clarification on their learning. Some example prompts are:
\begin{itemize}
    \item What is Big-$O$?
    \item What is a Linked List? Help me visualize a Linked List.
    \item What is a Left-Leaning Red-Black Tree? Give me an example of a Left-Leaning Red-Black Tree.
\end{itemize}

\subsubsection{Third Activity: Topic Heads-Up.} Students will request explanations from the GenAI tool regarding the forthcoming lab session's subject matter. To facilitate this phase, the instructor will give students a prompt to copy and paste into the tool directly. This step affords students a concise preliminary overview of the upcoming lab topic. The prompt's structure adheres to the \textit{``persona prompt pattern,''} \cite{white2023promptpattern} a strategy grounded in the observation that GenAI tends to yield enhanced and tailored outcomes when provided with a contextual role to assume.

We propose the following prompt template for students to run in the GenAI tool: \textit{``Act as a [top-level expert in the field you are trying to learn about, with qualifications or titles if necessary]. I want you to introduce me to [topic] briefly. I already know [list all relevant previous knowledge here]. The learning objectives for this section are as follows: [list of learning objectives to be satisfied].''}

We advocate for the active engagement of students in crafting personalized queries aligned with their chosen subject matter. An effective strategy involves leveraging Bloom's Revised Taxonomy \cite{bloomstax} to formulate precise learning objectives. Students can articulate these objectives in their vernacular and subsequently task a chat-based tool with rephrasing or rearticulating them employing taxonomy-specific terminology. As an illustrative exercise to unveil these capabilities, students can prompt ChatGPT to rephrase their self-defined learning objectives using the taxonomy. To facilitate this exercise, a designated prompt is suggested: \textit{``Act as a professor of engineering education to rephrase some learning objectives using Bloom's revised taxonomy. The learning objectives are: [List the learning objectives you want].''}

\subsection{In-Class Lab}

In this lab segment, students delve into the diverse applications of GenAI, understanding both its ideal use-cases and its limitations. An interactive component is woven into this phase, wherein students undertake the instructive role themselves to acquire mastery over a novel topic. To facilitate this interactive engagement, students must bring their laptops, participating in a synchronized learning experience guided by instructor-led demonstrations.

The session begins with a brief overview of GenAI, highlighting its essence and real-world applications. This is followed by a quick refresher on the basic prompts to ensure everyone is on the same page. The focus then shifts to GenAI's application in Education and Software Engineering. In the context of Education, discussions revolve around personalized examples, in-depth explanations, concept simplification, and generating practice questions. The Software Engineering segment delves into generating boilerplate code, crafting standard data structures and algorithms, identifying test cases, and aiding debugging. A key takeaway for students is recognizing GenAI's fallibility and building confidence in identifying its errors.

In light of these discussions, it becomes imperative for the instructor to guide students in the art of formulating effective prompts. The discussion could be as follows:

\subsubsection{First Step: Specific Context.} Each prompt should commence with role and context delineation. These two essential components constitute the prerequisites for eliciting high-quality outcomes, a principle encapsulated in the 'persona prompt pattern' \cite{white2023promptpattern}. For instance, consider the prompt: \textit{``Assume the role of a collegiate professor holding a Ph.D. in Computer Science, addressing a sophomore-level class on data structures and algorithms where students possess familiarity with Java.''} It is pivotal to consistently specify an expert's role alongside a specified level of proficiency or qualifications. The context element establishes the framework within which the response will be generated, tailored to the educational level of the students. Concluding the prompt by specifying the audience's familiarity with certain concepts enables the tool to utilize appropriate technical terminology and tailor the response to the student's existing knowledge base, thereby enhancing the effectiveness of the interaction.

\subsubsection{Second Step: Specific Request.}
Crafting effective prompts is pivotal for eliciting high-quality outcomes from GenAI interactions. While simple prompts often fall short of yielding optimal results, more elaborate and, at times, quantifiable instructions tend to enhance the quality of GenAI's outputs\cite{Zamfirescu23}. For instance, while the prompt \textit{``Give me a set of 10 problems for practicing Huffman encoding.''} may not guarantee a flawless set of problems, it is reasonable to anticipate a subset, typically 2-4 out of 10, will be good quality. It is crucial to acknowledge that perfection is not the norm. Emphasizing specificity in prompts, which often entails integrating multiple directives, can augment the depth and breadth of GenAI's responses. In this example:

\begin{itemize}
    \item \textit{``give me''} serves as a command, functioning as a specific and directed request aimed at the tool.
    \item \textit{``a set of 10 problems''} designates a precise numerical value to provide a range of options for selection, facilitating access to a focused collection of tasks (a problem set).
    \item \textit{``for practicing''} contributes to the overarching objectives. This element tailors the problems to support learning and active engagement with various aspects of the subject matter.
    \item \textit{``Huffman encoding''} encapsulates explicit terminology that pinpoints the exact subject under scrutiny. It is possible to delve into different levels of specificity, yielding varying degrees of success (typically greater success with heightened specificity). For instance, you could request questions about \textit{``compression trees''} in a broader sense, which would provide more theoretical inquiries. Conversely, soliciting problems concerning \textit{``using frequency analysis to create a Huffman Tree''} would furnish you with a set of scenarios dedicated explicitly to honing this particular skill.
\end{itemize}

\subsubsection{Third Step: Specific Goals.} The inclusion of goals within prompts is discretionary. For tasks that are concise and straightforward, the inclusion of goals may be optional. However, suppose the objective entails formulating an extensive series of questions, and examples, generating ideas, or addressing multifaceted tasks. Providing a broad outline of the intended direction is prudent in that case. This practice serves as a navigational aid for the GenAI tool, ensuring it aligns with the overarching purpose of the task at hand.

The instructor's domain expertise in the chosen topic plays a pivotal role in validating the accuracy of generated outcomes. Following the execution of each prompt, the instructor assumes the responsibility of corroborating the correctness of the result. This procedural step holds significance in aligning the conceptual underpinnings of the subject matter with the delivered outcome. To foster an engaging and interactive session, the instructor can effectively leverage discrepancies in the obtained results to encourage think/pair/share interactions \cite{felder2016teaching}, allowing students to showcase their grasp of the topic. As an illustrative example, consider the realm of Huffman Encoding. It is plausible that the GenAI tool may falter when tasked with constructing the tree, a scenario commonly observed when the algorithm merges two minimum frequency nodes into a shared parent node.

\subsection{Post-Lab}

The post-lab segment is a pivotal phase for reinforcing students' comprehension and offering instructors valuable insights into GenAI's integration within academic contexts. This phase encompasses a meticulously designed worksheet tailored to encompass a spectrum of tasks undertaken using GenAI. While the content need not be exclusively Computer Science-oriented, a worksheet section should be dedicated to the domain. This particular section should revolve around the subject matter covered in the class, encompassing activities such as problem-solving, question generation, critical evaluation of GenAI's topical outputs, or similar engagements.

A worksheet of this nature can be relatively straightforward, often comprising a single inquiry pertinent to the topic explored during the in-class lab, to be addressed as part of a subsequent homework assignment. This query should be designed in a two-fold manner: the initial segment prompting students to attempt problem-solving utilizing GenAI and the latter urging them to tackle the same issue via conventional conceptual approaches. From our practical observations, topics that expose GenAI's limitations can engender a pivotal juncture wherein students grapple with its shortcomings and then opt for a traditional problem-solving route. We advocate for instructors to engineer scenarios that tap into GenAI's technical capabilities. These scenarios can encompass diverse facets, such as implementing intricate details in a specific programming language, devising comprehensive test cases, identifying corner cases, establishing rational boundaries for asymptotic analysis, or seeking real-life applications.

Suppose the instructor opts to assign tasks that are not directly tied to the current topic covered in class. In that case, they can consider providing assignments that pertain to subjects of potential importance or relevance to the course yet may not be fully covered within the confines of traditional lectures. This approach could involve tasks prompting students to independently explore a subject that holds potential benefits but might be excluded from the list of core curriculum topics. For instance, tasks related to web development, unit testing, or other complementary areas could be incorporated to broaden students' skill sets and knowledge horizons.

A significant task related to the covered topic involves having students utilize GenAI to generate problem questions or examples about the subject they have learned:

\begin{itemize}
    \item Initiate the process by prompting them to devise questions regarding a familiar topic that has recently been taught and is within their comfort zone. Although it is possible to use the just-learned topic, generating practical learning objectives might pose a challenge. Providing learning objectives becomes optional if the topic is relatively straightforward and students have a solid understanding.
    \item Transition to having students generate questions about the topic introduced in the lab to strengthen their capacity for question generation and evaluation. In this scenario, learning objectives must be provided.
    \item Encourage students to select two appealing questions and elucidate their reasons for favoring them. Subsequently, instruct them to solve these chosen questions using GenAI and individually. Emphasize the value of employing GenAI after individual problem-solving, as it offers a distinct perspective, ensures topic comprehension, and enables comparison with GenAI-generated responses.
    \item Direct students to assess GenAI's answers based on the following criteria: technical accuracy, relevance to the posed question, and comprehensiveness of the response in addressing all aspects of the inquiry.
    \item Depending on the instructor's intended class discussion scope, there is an option to supply 1-5 instructor-generated questions related to the recently taught topic. Students would then solve these questions autonomously using GenAI and analyze the resultant answers. This task aims twofold: reinforce learning through independent application and critically evaluate GenAI's responses.
\end{itemize}

Concluding the cycle, reintroduce the survey from the initial stage of the Pre-Lab. This feedback mechanism aims to gauge the reception of GenAI utilization and provides valuable insights into students' evolving perspectives throughout the learning journey.

\section{Experience and Preliminary Data}

In the Summer of 2023, we implemented the AI-Lab framework within the context of CS256 - Data Structures and Algorithms, a pivotal undergraduate course offered by the Department of Computer Science at Anon University. We employed ChatGPT 3.5 as the chosen GenAI tool due to its open availability. Our decision to focus on Huffman Coding as the lab topic was twofold. 
Firstly, due to its alignment within the course, ensuring students possess a foundational grasp of the requisite underlying data structures. Secondly, Huffman Coding was chosen for its propensity to expose various errors in GenAI outputs during the coding algorithm execution, as evidenced by our experiences with ChatGPT 3.5.

We utilized the subsequent prompt to engage ChatGPT in introducing students to Huffman Coding: \textit{``Act as a \ul{computer science professor who has been trained how to teach well}. I want you to introduce me to \ul{Huffman Coding} briefly. I already know \ul{String pattern matching and Tries}. The learning objectives for this section are as follows: \ul{1. Analyze and illustrate the process by which Huffman coding compresses data. 2. Decompose and evaluate the structure and information presented in a Huffman tree. 3. Compute the bit requirement for a specific set of data using Huffman coding and validate its efficiency}.''}

We observed a notable attendance of 38 out of 41 students during the in-person session. The session's instructor meticulously navigated through the previously outlined pre-lab steps, acquainting the students with GenAI, ChatGPT, foundational prompts, and real-life instances. Progressing into the heart of the session, interactions between students and ChatGPT unfolded, centering on the definitions and significance of Huffman Coding. The instructor adeptly elucidated essential concepts as students engaged with the tool. To exemplify the algorithm's operation, students were prompted to collaborate in pairs, tasking ChatGPT with constructing Huffman Coding Trees for diverse strings. Swiftly, students were compelled to flag ChatGPT's errors—some of which teetered on the realm of absurdity—bringing them to the attention of their partners and instructors. By incorporating the Think/Pair/Share technique \cite{felder2016teaching} on ChatGPT's fallibility, students swiftly engaged in discourse and contrasted outcomes across pairs. Eventually, two groups (owing to logistical constraints) were designated to share their prompts and resultant outputs with the entire class. During this sharing, all groups adeptly discerned the errors, rectified them within the collective setting, and subsequently engaged in reflective discussions regarding the tool's accuracy.

The analysis of the anonymous pre-survey data unveiled intriguing insights. Firstly, at least 48.5\% of surveyed students were using GenAI tools for homework help. This figure might even be an underestimation, considering potential hesitations in self-reporting. Recent literature and news articles suggest that the actual number of students using GenAI and the variety of their methods might be more extensive than previously assumed by educators \cite{Yilmaz23, Terry23}. Furthermore, when examining the question about peer usage of GenAI on homework, it was revealed that at least 60\% of surveyed students were aware of fellow students incorporating GenAI tools into their coursework. To elaborate, 2.9\% of respondents indicated that 100\% of their peers employed GenAI, 17.1\% said 75\%-99\%, 20\%  said 50\%-74\%, 8.6\% said 25\%-49\%, 11.4\% said 1\%-24\%, 2.9\% said 0\%, and 37.1\% expressed ``uncertainty". Delving deeper into the pre-survey and post-survey findings, a discernible shift in students' perspectives concerning GenAI tool utilization for educational purposes emerged. Regarding the question on perception of GenAI reliability, the pre-survey results per option stood at 0\% for 1 (Very unreliable), 41.7\% for 2, 44.6\% for 3, 11.1\% for 4, and 2.8\% for 5 (Very reliable). Remarkably, the post-survey outcomes for the same question exhibited alterations: 10.3\% for 1 (Very unreliable), 27.6\% for 2, 41.4\% for 3, 13.8\% for 4, and 6.9\% for 5 (Very reliable). These variations underscore a perceptual shift in GenAI's reliability. Although derived from a controlled environment within a limited student population, these insights provide a glimpse of achievement in alignment with the proposed framework's objectives.

\section{Discussion and Future Work}

The rapid ascent of GenAI tools, mainly since November 2022, has sparked a wave of disruptions within computing academic circles, compelling instructors to recalibrate their strategies, adapt their approaches, or even adopt new stances in response to the evolving landscape \cite{LauGuo23}. While educators and researchers have anticipated the advent of AI tools in education, the integration of these diverse AI techniques into the intricate fabric of STEM education has posed formidable challenges, as underscored by Xu and Ouyang, who observed that \textit{``the application of AI technology in STEM education is confronted with the challenge of integrating diverse AI techniques in the complex STEM educational system''} \cite{Xu22}. Before the resounding impact of GenAI and its critical reception, AI tools were regarded as relevant but not yet transformative forces in education. As articulated by Zhai \textit{et al.}, \textit{``[Machine Learning] has transformed-—but not yet redefined—-conventional science assessment practice in terms of fundamental purpose, the nature of the science assessment, and the relevant assessment challenges''} \cite{Zhai20}.

Nonetheless, drawing from our academic roles and the firsthand experience we have gained with GenAI tools, we posit that GenAI's impact mirrors the transformative shifts witnessed with the advent of accessible platforms such as the Web, Search Engines, and the Internet within households. GenAI appears to be the latest addition to the toolkit available to the \textit{Net Generation} \cite{Jones2011}, a generation that includes students and instructors, each hailing from the same era of digital ubiquity. The allure of GenAI tools in the contemporary software development landscape is palpable, naturally enticing students to explore their potential. However, the pace of change within the academic realm does not match the rapid release of new GenAI tools in the tech industry. The inevitable consequence could be students appearing to excel in their coursework with apparent ease, giving rise to a situation where \textit{``our students have changed radically. Today's students are no longer the people our educational system was designed to teach''} \cite{Prensky05}. This dynamic poses intriguing questions about how educators must adapt their strategies to accommodate the paradigm shift facilitated by GenAI tools.

Given the recent emergence of readily accessible GenAI tools, the exploration of their application within academic environments is still in its early stages. A pivotal question surfaces within the context of integrating Generative AI into programming courses: \textit{``When instructing students in utilizing GenAI, what facets of pedagogy and subject matter remain within the purview of educators?''} We contend that, especially in courses emphasizing the foundational aspects of Computer Science, educators should prioritize cultivating mathematical reasoning and problem-solving skills over mere coding proficiency. Echoing the sentiments of Leslie Lamport, coding is akin to the mechanical act of typing in the realm of programming, a sentiment well-expressed in his observation, \textit{``coding is to programming what typing is to writing''} \cite{Lamport22}. This insight highlights the necessity of emphasizing broader cognitive skills that transcend specific programming languages or tools.

\begin{acks}
The authors would like to thank Purdue's Center for Instructional Excellence, at whose generative AI discussions we met and were encouraged to pursue our ideas. Furthermore, a special thanks to Emily Bonem and David Nelson for supporting and encouraging us initially and repeatedly as we execute our project.
\end{acks}

\newpage
\bibliographystyle{ACM-Reference-Format}
\bibliography{refs}

\appendix

\section{Pre/Post-Lab Surveys}
The following are the questions asked for students to answer as part of the activities of the pre/post-lab:

\subsection{Demographics}
\begin{enumerate}
    \item What academic year are you? Options: Checkmarks grid (Rows: High School, Undergrad, Graduate. Columns: 1st year, 2nd year, 3rd year, 4th year, 5th+ year).
    \item What is your major? Options: Computer Science, Data Science, Electrical Engineering (ECE or EE or CE or similar, High School (no major), Other.
    \item If not CS/DS, are you getting a CS minor? Options: Yes, No, N/A.
    \item How many programming courses have you taken before now (not inclusive of current courses)? Options: 0, 1-2, 3-4, 5+.
    \item Have you had any professional experiences where programming is required (e.g., internships, paid coding jobs, etc.)? Options: Yes, No.
    \item How much have you attempted to use any generative AI tool (for recreational/academic/professional purposes)? Options: Not at all, Played with it only a couple of times to learn about it, Tried to use it a few times, Use it semi-regularly, Use it frequently.
\end{enumerate}

\subsection{Usage}
\begin{enumerate}
    \item How many of your peers in CS and Engineering programs are using GenAI for homework help? Options: 100\%, 75\%-99\%, 50\%-74\%, 25\%-49\%, 1\%-24\%, 0\%, No idea.
    \item How frequently do you use GenAI to get help with \_\_\_\_\_? Options: Checkmarks grid (Rows: Conceptual Questions, Debugging, Homework Problems, Programming Projects. Columns: Very frequently, moderately frequent, on occasion, never).
\end{enumerate}

\subsection{Perception}
\begin{enumerate}
    \item In general, how open are you to GenAI being allowed in a collegiate class? Options: Scale from 1 (Very opposed to using it) to 5 (Very open to use it).
    \item How open are you to using GenAI to get help with \_\_\_\_\_? Options: Checkmarks grid (Rows: Conceptual Questions, Debugging, Homework Problems. Columns: Very uncomfortable, somewhat uncomfortable, neutral, somewhat comfortable, very comfortable).
    \item How comfortable are you to using GenAI to get help with \_\_\_\_\_? Options: Checkmarks grid (Rows: Conceptual Questions, Debugging, Homework Problems. Columns: Very uncomfortable, somewhat uncomfortable, neutral, somewhat comfortable, very comfortable).
    \item How reliable do you think GenAI is for most problems you have in programming classes? Options: Scale from 1 (Very unreliable) to 5 (Very reliable).
    \item What potential benefits do you see in using GenAI to assist students during a course? Option: Short written answer.
    \item What potential challenges do you see in using GenAI to assist students during a course? Option: Short written answer.
\end{enumerate}

\end{document}